\documentclass[11pt,letterpaper]{article} 
\usepackage[margin=1in]{geometry}
\usepackage{graphicx}
\usepackage{amssymb}

\usepackage{times}

\usepackage{url}
\usepackage[sort]{cite}
\usepackage{comment}
\usepackage{amsmath}
\usepackage{array}
\usepackage{balance}

\newcommand{\Sign}{\ensuremath{\mathcal{S}}}
\newcommand{\PK}{\ensuremath{\widehat{PK}}}

\newcommand{\Qu}{\textsf{Q}}

\newcommand{\key}{\textsf{key}}

\newcommand{\RCPI}{\textsf{RCP-I}}
\newcommand{\RCPII}{\textsf{RCP-II}}
\newcommand{\RCPqpI}{\textsf{RCPqp-I}}
\newcommand{\RCPqpII}{\textsf{RCPqp-II}}
\newcommand{\OTPK}{PK} \newcommand{\OTG}{\ensuremath{G}}
\newcommand{\RT}{\ensuremath{\mathcal{RT}}}

\newcommand{\ooom}[1]{$1$-out-of-\ensuremath{#1}}

\newcommand{\Z}{\ensuremath{\mathbb{Z}}}
\newcommand{\ZOTG}{\ensuremath{\Z_{|\OTG|}}}

\long\def\symbolfootnote[#1]#2{\begingroup\def\thefootnote{\fnsymbol{footnote}}\footnote[#1]{#2}\endgroup}
\title{Adding Query Privacy to Robust DHTs}

\author{Michael Backes$^{1,2}$ \hspace{0.5cm} Ian Goldberg$^{3}$ \hspace{0.5cm} Aniket Kate$^{1}$ \hspace{0.5cm}Tomas Toft$^{4}$\\
$~$\\
\small{$~^{1}$Max Planck Institute for Software Systems (MPI-SWS), Germany}\\
\small{$~^{2}$Saarland University, Germany}\\
\small{$~^{3}$University of Waterloo, Canada}\\
\small{$~^{4}$Aarhus University, Denmark}
}
\date{ }

\begin{document}
\maketitle

\begin{abstract}
Interest in anonymous communication over distributed hash tables
(DHTs) has increased in recent years. 
However, almost all known solutions solely aim at achieving sender or
requestor anonymity in DHT queries.  In many application scenarios, it is crucial that the queried key remains secret
from intermediate peers that (help to) route the queries towards their
destinations. In this paper, we satisfy this requirement by presenting
an approach for providing privacy for the keys in DHT queries.

We use the concept of oblivious transfer (OT) in communication over
DHTs to preserve query privacy without compromising 
spam resistance.  Although our OT-based approach can
work over any DHT, we concentrate on communication over {\em robust}
DHTs that can tolerate Byzantine faults and resist spam.  We
choose the best-known robust DHT construction, 
and  employ an efficient OT protocol well-suited for
achieving our goal of obtaining query privacy over robust DHTs.
Finally, we compare the performance of our privacy-preserving
protocols with their more privacy-invasive counterparts. We observe
that there is no increase in the message complexity and only a small
overhead in the computational complexity.

\end{abstract}

\section{Introduction}\label{sec:Intro}
In the digital society, our online activities are persistently
recorded, aggregated, and analyzed. Although worldwide electronic data
privacy laws and organizations such as EFF \cite{eff} and EPIC~\cite{epic} try to challenge
this pervasive surveillance through policies and protests, privacy
enhancing technologies (PETs) are key components for establishing a
suitable privacy protection mechanism from the technology side. 
The interest in developing novel PETs is increasing
for a variety of reasons, ranging from the desire to share and access
copyrighted information without revealing one's network identity, to
scalable anonymous web
browsing~\cite{salsa:nambiar, MTHK09, mittal:shadow, PRR09, WMB10}.
In this paper, we study privacy in the
peer-to-peer (P2P) paradigm, 
a popular approach to providing large-scale decentralized services.

In the P2P paradigm, distributed hash tables
(DHTs)~\cite{ratnasamy:scalable,stoica_etal:chord,RD'01,ZHS+'04} are
the most common methodology for implementing structured routing.  
Similar to hash tables, a DHT is a data structure that efficiently maps {\em
keys} onto {\em values} that are stored over a distributed overlay
network. 
However, unlike hash tables, DHTs can scale to extremely large number of
key-value pairs as the mapping from keys to values is distributed among all peers.
In order to obtain a value associated with a key, a
requester (a sender) in a DHT routes the key through a small fraction of
the network to reach the receiver that has stored the value.
DHTs can also handle continual arrivals and departures of peers, and small-scale
modifications to the set of peers do not disturb the mapping  from keys to values significantly.

In a DHT, privacy may be expected for the sender, the receiver or the queried key.
Ensuring the anonymity of
senders and requesters in DHTs has received considerable attention in
the privacy community~\cite{salsa:nambiar,MTHK09,mittal:shadow,PRR09,WMB10}.
Privacy of the queries / keys, i.e., keeping the keys
secret from intermediate peers that route the queries towards their
destinations, is also equally important: in many scenarios such as censorship resistance, this query
privacy constitutes a necessary condition for sender and
requestor anonymity.  In this paper, we present a practical approach
to obtain privacy for queries in {\em robust and spam-resistant} DHTs where a fraction of peers
may behave maliciously.

\subsection{Contributions}
Almost all anonymity solutions for DHTs
try to provide anonymity to a sender or a requester in a DHT lookup, upload or request.
It may also be necessary that the queried key remains secret from peers
that route the corresponding requests in some situations. We call this property {\em query privacy}.
In this case, an intermediate routing peer should be able to suggest a next peer or a set of next peers
without determining the key being searched for.
Example application scenarios for this property can be
protection against mass surveillance or censorship, preventing tracking and data-mining activities on users requests,
and providing opportunities to access material that is socially deplored, embarrassing or problematic in society.

\emph{Recursive} routing and \emph{iterative} routing are the two approaches to route information in DHTs.
In the \emph{recursive} routing approach,
obtaining query privacy looks infeasible, if not impossible.
This results from the fact that the intermediate router itself decides to which peer 
to forward a request. Assuming that every peer
is under the control of an individual, it is always possible for the controller to figure out the next peer 
for the request.
On the other hand, query privacy in \emph{iterative} routing, 
which is also a commonly used routing approach over robust DHTs, can be trivially obtained if
every peer sends its complete routing table to the requesting peer.
The requester can then determine the next peer itself and send the
request.  However, this solution may make it significantly easier to mount
spamming attacks in the systems: a malicious sending peer can easily
gather a significant amount of routing information, and use it to
determine and target peers that hold specific keys.

We instead use the oblivious transfer (OT) primitive~\cite{rabin81,EGL85}. Given a peer holding a database,
OT allows a requester with a key to obtain a database entry associated with the key, such that
the requester does not get any information about other database entries
and the database owner does not learn the requester's key.
Therefore, OT perfectly fits our requirements of obtaining query privacy without divulging additional routing information.
We use the OT protocol by Naor and Pinkas~\cite{NP01}
in the best-known robust DHT constructions by Young et al.\ \cite{young:practical, ToN}
(their RCP-I and RCP-II protocols)
to obtain our goal of query privacy in robust DHTs.
Importantly, our query privacy mechanism does not increase
the message complexity of the RCP-I and RCP-II protocols
and an increase in the computation cost is also not significant.
We elaborate on our exact choice of OT protocol in Section~\ref{sec:OT},
and discuss robust DHT constructions in Section~\ref{sec:Related}.

The employed OT protocol~\cite{NP01} is a simple indexed OT protocol,
where the database contains index-value pairs and a requester inputs an index.
However, a query for a routing table entry is an interval membership (or a range) query and not an index query.
Therefore, to prevent a requesting peer from obtaining {\it any} additional information,
we could have employed the concept of conditional OT (COT)~\cite{DOR99} and
used the interval-membership strong COT (I-SCOT) protocol by Blake and Kolesnikov~\cite{BK09}.
However, the I-SCOT protocol is expensive in terms of both computation and communication.
We observe that by releasing the upper and lower bounds of a routing table entirely to the requester,
large improvements can be made. There is no information in these range boundary values for a malicious requesters
in terms of spamming as they do not convey any information regarding the identities of the key owners.
However, given this information, the requester {\it will
know} the desired entry (index), allowing the use of OT instead of the more complex COT.

Private information retrieval (PIR)~\cite{CKGS98} is a weaker form of OT,
where more information may be revealed than asked for;
e.g., sending the complete routing table is a trivial PIR protocol.
PIR protocols can be less costly  than OT in terms of computation,
but the risk of spamming persists with all non-OT PIR protocols, and hence we avoid them.

\paragraph{Outline}
The rest of the paper is organized as follows. In Section~\ref{sec:Related}, we survey the literature on robust DHTs.
Section~\ref{sec:Prelim} describes our system model, 
while Section~\ref{sec:Tools} overviews the cryptographic tools used in our constructions.
In Section~\ref{sec:schemes}, we present the robust communication protocols that preserve query privacy.
In Section~\ref{sec:System}, we analyze and discuss performance and systems issues.
Finally, we conclude in Section~\ref{sec:Con}.
We include a detailed description of the employed OT protocol in Appendix~\ref{app:OT}.

\section{Background and Related Work}\label{sec:Related}
Malicious behaviour is now common over the Internet.
Lack of admission control mechanisms in DHT systems make them particularly vulnerable to these malicious (or Byzantine) attacks~\cite{sit:security,wallach:survey}.
Such attacks can not only pollute the data that is available over DHTs~\cite{liang:pollution}, 
but also poison the indices by creating fake data identifiers~\cite{liang:poison}.
They may further create Sybil identities and disrupt communication between well-behaving peers by spamming.
The concern is quite serious, since large-scale P2P systems in existence today
(e.g., Azureus or Vuze DHT~\cite{falkner:profiling} or KAD DHT~\cite{steiner:global})
see millions of users every day. Along with the basic file sharing application, 
there are proposals for using P2P systems to re-implement the Domain Name System~\cite{pappas:comparative},
mitigate the impact of computer worms~\cite{aspnes:worm} and protect archived data~\cite{Vanish,Unvanish}.
These applications would benefit from tolerance against Byzantine behaviors.
As a result, a number of solutions have been defined that can provably tolerate Byzantine faults over P2P systems 
(e.g., \cite{awerbuch:towards,awerbuch:towards2,awerbuch:random,saia:reducing,
fiat:making,naor_wieder:a_simple,HK,johansen:fireflies,young:practical}).
Due to the popularity of DHTs, the majority of these solutions are built to work over DHTs 
and the resulting constructions are called {\it robust} DHTs.

In robust DHTs, malicious attacks are generally dealt with using the concept of {\em quorums}~\cite{HK,naor_wieder:a_simple,saia:reducing,awerbuch_scheideler:group,fiat:making,awerbuch:towards,awerbuch:towards2,awerbuch:random}.
 A quorum is a set of peers such that a minority of the members suffer adversarial faults. 
Typically, it consists of $\Theta(\log n)$ nodes where $n$ is the total number of nodes in the underlying DHT.
A DHT quorum replaces an individual DHT peer as the atomic unit and malicious behaviour by an adversary is overcome by majority action;
e.g., the content may be stored in a distributed and redundant fashion across members of a quorum
such that it cannot be polluted by a small fraction of host peers. 
Poisoning attacks can be mitigated by having peers belonging to the same quorum validate routing information before it is advertised. 
If a peer violates the protocol, then it is possible to remove it from the quorum, which effectively removes them from the system.

\begin{figure*}[ht]
\begin{center}
\begin{tabular}{p {2.5in} p {0.5in} p {2.5in}} 
\includegraphics[scale=0.2]{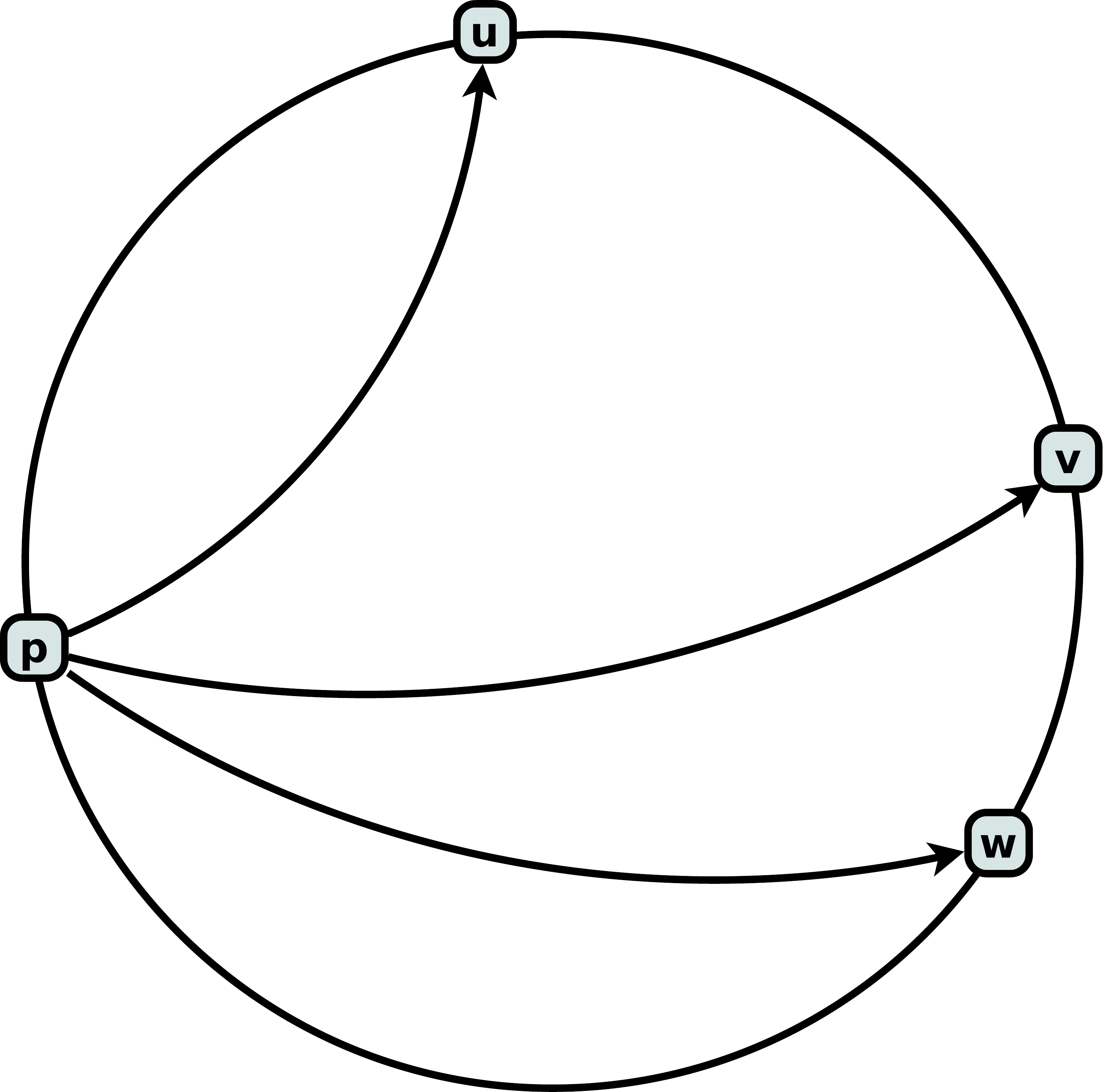} & &\includegraphics[scale=0.2]{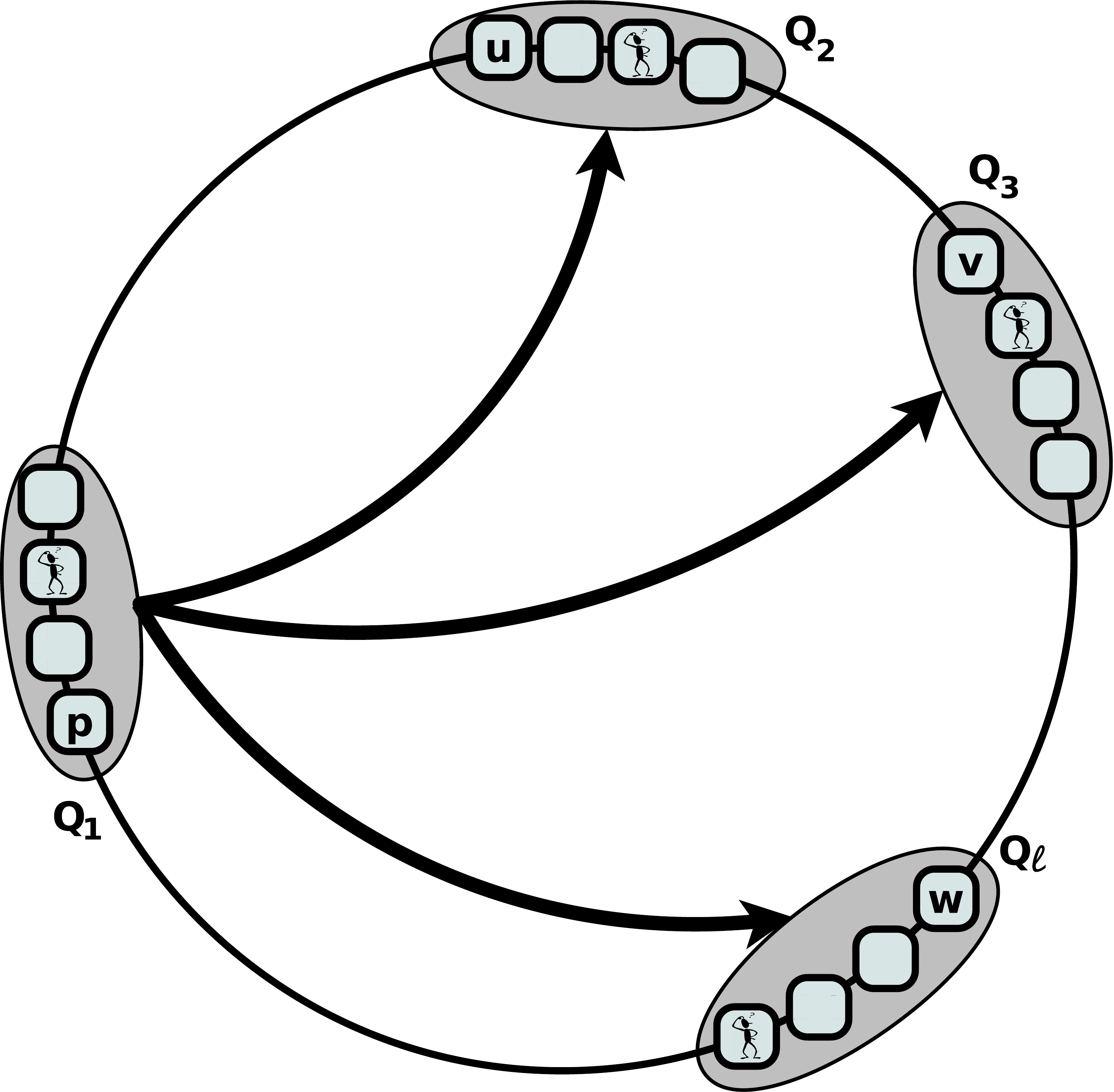}\\
\end{tabular}
\end{center}
\small{Peers $u$, $v$, and $w$ are in the routing table (\RT) of peer $p$ on a DHT. Correspondingly, in an quorum topology with $p\in{\Qu_1}$, $u\in{\Qu_2}$, $v\in{\Qu_3}$ and $w\in{\Qu_\ell}$, quorums $\Qu_2$, $\Qu_3$, and $\Qu_\ell$ are linked to quorum $\Qu_1$ in its \RT.
Thick lines signify inter-quorum links. Each quorum has size $\eta=\Omega(\log{n})$ and must have strictly fewer than $1/3$ faulty peers.}
\caption{Quorum Topology over DHTs
\label{fig:Quorum}}
\end{figure*}

Protocols using quorums are split between those that use iterative and those that use recursive approaches.
When sending a request using the recursive approach, a sending peer contributes one message (its request), 
while its DHT has to generate $O(\log{n})$ messages.
In the iterative approach, a sending peer has to contribute an equal number of message as its DHT.
Consequently, while dealing with Byzantine faults, 
the iterative approach is more common than the recursive approach as
the former provides better protection against the spamming attack than the latter.

The common way such quorums are utilized is as follows: a request $m$
originating from a peer $p$ traverses a sequence of quorums $\Qu_1$,
$\Qu_2$, $\dots$, $\Qu_{\ell}$ until a destination peer is reached. A
typical example is a query for content where the destination is a peer $q$
holding a data item. Initially $p$ notifies its own quorum $\Qu_1$ that it
wishes to transmit $m$. Each peer in $\Qu_1$ forwards $m$ to all peers in
$\Qu_2$. Every peer in $\Qu_2$ determines the correct message by majority
filtering on all incoming messages and, in turn, forwards it to all peers in the
next quorum. This forwarding process continues until the quorum
$\Qu_{\ell}$ holding $q$ is reached.  Assuming a majority of correct peers
in each quorum, transmission of $m$ is guaranteed.  

Unfortunately, this simple protocol is costly. If all quorums have size $\eta$ and the path length
is $\ell$, then the message complexity is $\ell \eta^{2}$. Typically, for a
DHT of $n$ nodes, $\eta=\Theta(\log{n})$ and, as in Chord~\cite{stoica_etal:chord}, $\ell=O(\log{n})$, which gives
$O(\log^{3}n)$ messages; this is likely too costly for practical values of $n$.
Saia and Young~\cite{saia:reducing} mitigate this problem using a randomized protocol
which provably achieves $O(\log^2{n})$ messages in expectation; 
however, the constants in their protocols are prohibitively large.

Recently, Young et al.\ \cite{young:practical} demonstrated that the problem can be solved using threshold cryptography~\cite{Des94}.
Using a distributed key generation (DKG) protocol over the Internet~\cite{kate:distributed} and a threshold digital signature scheme~\cite{Bol03}, 
they design two robust communication protocols, RCP-I and RCP-II, that respectively require $O(\log^2{n})$ messages 
and $O(\log{n})$ messages in expectation.
Importantly, these protocols can tolerate adversarial peers up to 
any number less than $1/3$ of a quorum in the asynchronous communication setting and 
less than $1/2$ of a quorum in the synchronous communication setting.
They also do not require any trusted party or costly global updating of public/private keys outside of each quorum.
The protocols work in the elliptic curve cryptography (ECC) based discrete logarithm setting, and
its security is based the gap Diffie--Hellman (GDH) assumption~\cite{BLS01}.
The paper also includes
results from microbenchmarks conducted over PlanetLab 
showing that these protocols are practical for deployment under significant levels of churn and adversarial behaviour.
We find this work to be the most up-to-date solution for robust and 
spamming-resistant communication in DHTs and use it as a starting point towards query privacy.

Privacy in communication over DHTs has also been under consideration over the last few years \cite{salsa:nambiar,MTHK09,mittal:shadow,PRR09,NPMA'10,WMB10}. 
However, most of these PETs concentrate on sender (or requester) privacy,
and generally aim at a scalable anonymous web browsing system: a future replacement for Tor~\cite{Tor}.
Our aim in this paper is different; we want to achieve privacy for keys in DHT queries (or query privacy). 
Nevertheless, we observe that
our query privacy mechanism can further enhance anonymity in almost all of the above PETs. 
Our approach is also significantly better in terms of message complexity than redundant routing~\cite{NPMA'10},
where a requester makes multiple queries to confuse an observer.

\section{System Model and Assumptions}\label{sec:Prelim}
In this section, we discuss the quorum-based DHT system model,
and the adversary and communication assumptions that we make in our protocols.
As we develop our anonymity solution on top of robust communication protocols 
by Young et al.\ \cite{young:practical}, our model is nearly the same as their model.

For ease of exposition, we do not consider the link failures and crash-recovery mechanism
in that work, which in turn follows from the underlying DKG architecture~\cite{kate:distributed}.
However, our protocols indeed work even under these assumptions without any modification.

\subsection{Adversary and Communication Assumptions}\label{sec:Assump}
We work in the asynchronous communication model (unbounded message delays) with Byzantine faults.
However, to ensure the liveness of the protocols, we need the {\em weak synchrony} communication assumption by Castro and Liskov~\cite{castro:practical}, 
which states that the message delay does not grow longer indefinitely.
Note that this assumption arrives from the underlying robust communication protocols 
and is unrelated to our privacy preserving mechanism.

In a P2P system, each peer is assumed to have a unique name or identifier $p$ and an IP address $p_{\mbox{\tiny addr}}$. 
Peers $p$ and $q$ can communicate directly if each has the other in its routing table (\RT).

Similar to the majority of anonymous communication networks~\cite{Tor,MTHK09,mittal:shadow,PRR09},
we do not assume a global adversary that can control the whole network and break
anonymity by observing all communication by every peer. 
Such an adversary seems impractical in large-scale geographically distributed DHT deployments.
However, we assume that our partial adversary knows the network topology
and controls a small fraction of the DHT peers.
Following prior works~\cite{rodrigues:rosebud, reconfigure, SLL10, young:practical, ToN}, 
we consider around $10\%$ of all peers to be under adversarial control.
The adversary cannot observe communication at the majority of nodes;
however, it may try to break query privacy, spam honest nodes, 
or disrupt the communication by actively attacking traffic
that reaches peers under its control.

We assume that the $10\%$ adversarially controlled nodes are spread out evenly over the DHT,
and strictly less than $1/3$ of the peers in any quorum are faulty
which is the best possible resiliency in the asynchronous setting.
This bound on the adversary is possible
using mechanisms like the {\em cuckoo-rule} developed by Awerbuch
and Scheideler~\cite{awerbuch:towards}, which restricts the adversary from
acquiring many peers in the same quorum.
Further, all faulty peers in a quorum may be under the control of a single adversary,
and collude and coordinate their attacks on privacy, safety and liveness.

Finally, our adversary is computationally bounded with
security  parameter $\kappa$. We assume that it is infeasible for the adversary to solve 
the GDH problem~\cite{BLS01} in an appropriate setting for signatures and the
decisional Diffie--Hellman (DDH) problem~\cite{Bon98} in another setting for OT.

\begin{figure}[ht]
\begin{center}
\includegraphics[scale=0.39]{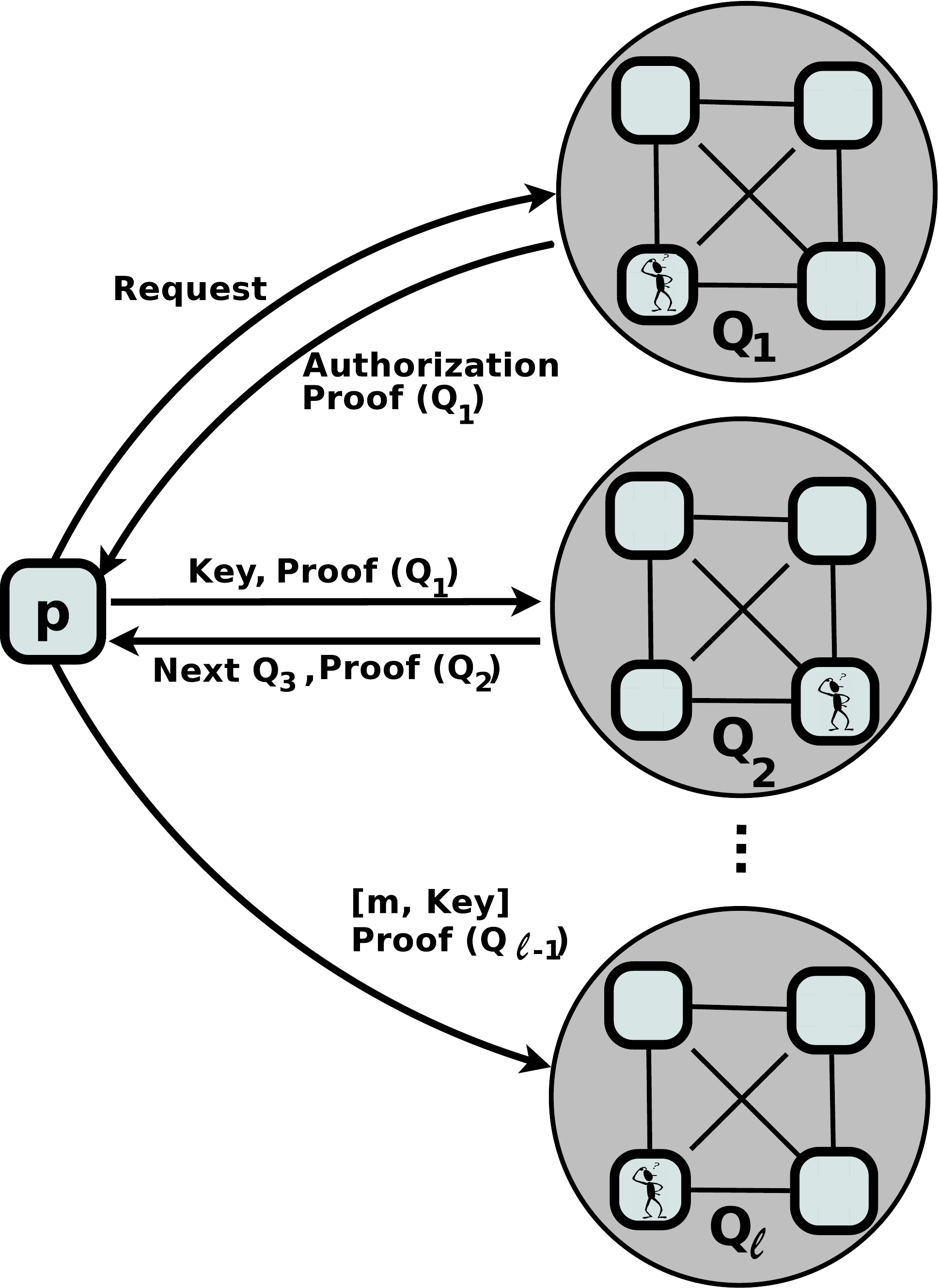}
\end{center}
\small{A peer $p$ sequentially communicates with $\Qu_1$, $\Qu_2$, and so on, until it reaches $\Qu_\ell$
who owns the searched-for \key.}
\caption{Iterative Communication in Robust DHTs using Quorums
\label{fig:overview}}
\end{figure}

\subsection{Quorums}\label{sec:Quorum}
In a variety of approaches used to maintain quorums,
one may view the setup of quorums as a graph where peers correspond to quorums and 
edges correspond to communication capability between quorums. This is referred to as the {\it
  quorum topology} in the literature. Figure~\ref{fig:Quorum} shows how quorums can be linked 
 in a DHT such as Chord~\cite{stoica_etal:chord}.

We assume the following four standard invariants~\cite{young:practical}
are true for the quorum topology under consideration:
\begin{description}
\item[Goodness.] Each quorum has size $\eta=\Omega(\log{n})$ and must have strictly fewer than $1/3$ faulty peers.
\item[Membership.] Every peer belongs to at least one quorum.
\item[Intra-Quorum Communication.] Every peer can communicate directly to
  all other members of its quorums.
\item[Inter-Quorum Communication.] if $\Qu_i$ and $\Qu_j$ share an edge in
  the quorum topology, then $p \in \Qu_i$ may communicate directly with any
  member of $\Qu_j$ and {\it vice versa}.
\end{description}

To the best of our knowledge, no practical implementation of a quorum
topology yet exists.
However, as indicated in the literature~\cite{awerbuch:random,awerbuch:towards,awerbuch:towards2,fiat:making,naor_wieder:a_simple}, 
maintaining the above four invariants looks plausible in real-world DHTs.

In a DHT where the above four invariants are maintained,
the general communication mechanism in Young et al.\ \cite{young:practical} works as shown in Figure~\ref{fig:overview}.
Assume that a peer $p$ wants to send a query $m$ associated with a \key\ that belongs to quorum $\Qu_\ell$,
which it does not know.  The recipients of the request are generally a set of peers $D \subseteq \Qu_\ell$.
 Peer $p$ requests authorization from peers in its quorum $\Qu_1$.
These authorizations are based on a {\it rule set}~\cite{fiat:making} that defines acceptable behavior in a
quorum (e.g., the number of data lookup operations a peer may execute during a predefined time period).
This rule set is known to every peer within a quorum and is possibly the same across all quorums;
it reduces the impact of spamming attacks.
Peer $p$ receives \textsc{Proof}($\Qu_1$) in the form of a signature if authorized. 
It then sends this to quorum $\Qu_2$ from its routing table, which is
responsible for the \key\ being searched for.
One or more members of $\Qu_2$ verify the signature and provides $p$ routing information and a \textsc{Proof}($\Qu_2$) for $\Qu_3$, which
will convince $\Qu_3$ that $p$'s actions are legitimate (i.e., approved by its quorum). 
The protocol continues until $p$ reach $\Qu_{\ell}$.

As mentioned in Section~\ref{sec:Related}, it possible to achieve robust communication 
without using any of the above cryptography.
However, use of cryptography provides efficiency and reduce the message complexity by at least a linear factor. 
Note that we do not discuss membership update operations for quorums in this paper
as they remain exactly the same as those in previous work~\cite{young:practical, ToN}.

\section{Cryptographic Tools}\label{sec:Tools}
Here, we describe the cryptographic tools that we use in our solution.
In particular, we review distributed key generation, threshold signature and oblivious transfer protocols.

 \subsection{Threshold Signatures}\label{sec:Sign}
The use of distributed key generation (DKG) and threshold signatures in our privacy preserving schemes
comes from the underlying robust DHT architecture.
In this architecture, threshold signatures are used to authenticate the communication between quorums. 
In an $(\eta,t)$-threshold signature scheme, 
a signing (private) key $sk$ is distributed among $\eta$ peers either by a trusted dealer (using verifiable secret sharing)
or in a dealerless fashion (using DKG).
Along with private key shares $sk_i$ for each party, the distribution algorithm also generates a verification (public)
key $PK$ and the associated public key shares $\PK$.  
To sign a message $m$, any subset of $t+1$ or more peers use their shares to generate the
signature shares $\sigma_i$.  Any party can combine these signature shares
to form a message-signature pair $\Sign = (m,\sigma) = [m]_{sk}$ that can be
verified using the public key $PK$.

In this work, we refer to a message-signature pair $\Sign$ as a signature.
Further, it is possible to verify the individual signature shares $\sigma_i$ using the public key shares $\PK$.
We assume that no computationally bounded adversary that 
corrupts up to $t$ peers can forge a signature $\Sign' = (m',\sigma')$ for a message $m'$.
Malicious behaviour by up to $t$ peers cannot prevent generation of a signature.

Among three known practical threshold signature schemes \cite{GJKR96,Sho00,Bol03}, Young et al.\ employed
the threshold version~\cite{Bol03} of the Boneh-Lynn-Shacham (BLS) signature scheme~\cite{BLS01} for their robust DHT design.
They reason that, unlike Shoup's construction~\cite{Sho00}, the key generation in threshold BLS signature scheme does not mandate a trusted dealer,
and unlike Gennaro et al.'s construction~\cite{GJKR96}, the signing protocol in threshold BLS signature scheme does not require any interaction among peers
or any zero-knowledge proofs. 
They also mention efficiency of the BLS signature scheme in terms of
size and generation algorithm as compared to the other options
and employ it
to authenticate the communication between the quorums.

\subsection{Distributed Key Generation---DKG}\label{sec:DKG}
As a trusted party is not feasible in the P2P paradigm, 
the underlying robust DHT architecture also needs a complete distributed setup 
in the form of DKG to generate distributed signing keys.
An $(\eta,t)$-DKG protocol allows a set of $\eta$ nodes to construct a shared secret key $sk$ such that 
its shares $sk_i$ are distributed across the nodes and no coalition of fewer than $t$ nodes may reconstruct the secret.
In the discrete logarithm setting, 
there is also an associated public key $PK$ and a set of public key shares $\PK$ in DKG 
for verification as required for threshold signatures.

For the robust DHT architecture, Young et al.\ use a DKG protocol~\cite{kate:distributed}
defined for use over the Internet. We continue to use threshold BLS signatures over this DKG setup in
our privacy preserving enhancement.

\subsection{Oblivious Transfer---OT}\label{sec:OT}
The first notion of oblivious transfers was introduced in 1981 by Rabin
\cite{rabin81}. A \ooom{2} oblivious transfer (OT)~\cite{EGL85} allows a
chooser\footnote{This is sometimes denoted ``receiver'' in the OT
  literature; we use the term ``chooser'' to avoid confusion with the
  overall receiver of message $m$ in the surrounding DHT protocol.}
$p$ to decide between two messages held by a server\footnote{This is
  typically denoted ``sender'' in the OT literature; we use the
  term ``server'' to avoid confusion with the overall sender of
  message $m$ in the surrounding DHT protocol.}  $q$. Moreover, OT
protocols also guarantee that the server learns nothing, while the
chooser obtains at most one of the messages. The concept may be
generalized to \ooom{\nu} OT, where $q$ holds $\nu$ messages from
which $p$ may pick only one.
In this work we will use this to obtain the relevant entry of the
routing table from a quorum; the use of oblivious transfers ensures
that the query remains secret, while at the same time spamming is
prevented, since a malicious $p$ is guaranteed to receive only a single
entry.

We utilize an OT protocol by Naor and Pinkas~\cite[Protocol~3.1]{NP01} as
it fulfills all our needs; see Appendix~\ref{app:OT} for an overview.
The protocol provides \ooom{\nu} string OT, as we require.
It is round optimal and requires only one message per
party ({\it OT-request} from $p$ and {\it OT-response} from $q$), 
except an {\it OT-setup} message that we may piggyback in the
surrounding protocol.
Moreover, it requires no zero-knowledge proofs, and 
also works in the elliptic curve cryptography (ECC) setting.
The computation complexity of the protocol is dominated by the number of exponentiations;
both server and chooser must on average perform two of these.
In addition to the low computational costs, the overall communication amounts to roughly $3\nu$ group elements.

The construction of Naor and Pinkas allows transfer of group elements; i.e.,
strings of approximately 256 bits in the ECC setting. This is not sufficient for an
entire entry of a routing table. Rather than increasing the group
size or performing multiple OTs, we simply
let a peer $q$ symmetrically (AES) encrypt each entry of the routing table using a
random key.
The encrypted table is then sent to peer $p$ who uses an OT
execution to obtain the AES key for the relevant entry from peer $q$.

For protocol \RCPqpI\ in Section~\ref{sec:RCP-I}, we will require a chooser $p$ to run an OT with multiple
members of the same quorum. We could reduce $p$'s computation by 
ensuring that all parallel OT instances are verbatim copies here. 
This would naturally require that the all servers use the same source of randomness
for {\it OT-setup} and for AES keys.
This can be achieved easily using a parameterized pseudorandom function (PRF): $\phi(r,\cdot)$.
The private key $r$ required for $\phi$ can easily agreed upon as part of a DKG execution,
as it should be known to all quorum members.
When the quorum executes an OT instance with chooser $p$, it may use $p$'s 
message itself as an input to PRF $\phi$.
This PRF-based modification does not have any effect on the OT security proof
as all parallel OT instances are verbatim copies.

\paragraph{Other Possibilities} A natural question to ask is whether OT is really required, or whether
another protocol could achieve the desired goal more efficiently. 
Although PIR protocols appear to be an alternative, 
they are not an acceptable alternative because they leak routing information.
Further, 
computational PIR protocols have similar cost as the selected OT protocol~\cite{NP01}.
For that matter, most non-trivial PIR is essentially OT as well.

Theoretically better OT protocols also exist, e.g.~Lipmaa's OT
protocol\cite{Lipmaa05}, which provides \ooom{\nu} OT with $O(\log^2 \nu)$
multiplicative overhead on communication (of a single entry). 
For the proposed protocol by Naor and Pinkas, the overhead is linear 
which, in theory, is clearly worse.
However, our approach is better in the present setting when we consider numbers from real-world DHTs.
 With more than million peers in a practical DHT, we will have $\nu\approx 20$. 
For $\nu\approx 20$, $\log^2 \nu \approx 20$ and linear communication without any hidden constant is quite acceptable.
A generic \ooom{2} OT protocol of Peikert et~al.~\cite{PVW08} requires only two messages, 
and roughly five exponentiations per party. However, this is still more than the
amortized cost of the \ooom{\nu} OT of Naor and Pinkas and we do not use it.
While we cannot rule out the possibility of a more efficient protocol,
it seems highly unlikely.



Finally, hiding the range values in routing table entries seems possible, 
but it is most likely infeasible in practice. Blake and Kolesnikov~\cite{BK09} provides a
\ooom{2} conditional OT (COT) based on the greater-than
relation. Their protocol has a blowup of a factor linear in the
bitlength of the key. This blowup is needed in order to compute the
greater-than relation. In addition to this, there are two critical
issues that must be solved before COT can be used for hiding the range values:
1) the present work~\cite{BK09} requires a \ooom{\nu}
conditional oblivious transfer 2) the
protocols of \cite{BK09} are only secure against semi-honest
adversaries. Neither seems impossible to solve, but both
appear to incur a significant blowup. 
Nevertheless, as no routing information is lost through range boundaries, 
we need not consider these.

\section{Adding Query Privacy}\label{sec:schemes}
Young et al.\ \cite{young:practical} present two robust communication protocols 
using quorums and threshold cryptography:
{\sf RCP-I} and {\sf  RCP-II}. As described in Section~\ref{sec:Quorum},
both these protocols work in the general communication architecture shown in Figure~\ref{fig:overview}.
They use threshold BLS signatures over the DKG architecture explained in sections \ref{sec:Sign} and \ref{sec:DKG}.
In this section, we provide query privacy to the above protocols using the
OT primitive explained in Section~\ref{sec:OT} to define protocols {\sf RCPqp-I} and {\sf RCPqp-II}.

\subsection{System Setup}
We start our discussion by describing the setup required for our protocols. 
For clarity of description, we also briefly review routing tables (\RT) in quorum-based DHTs.

\begin{description}
\item[Initiation.] Before the system becomes functional, 
the initiator has to choose appropriate groups and other setup parameters for the BLS signature and OT protocols.
Note that there are no trust assumptions required during this step, as these parameters can be selected
from the well-known standards.
\item[Distributed Key Generation.] A DKG instance is executed, when a quorum gets formed in DHTs. 
At the end of an execution, each quorum $\Qu_i$ is associated with a (distributed) public/private key pair $(PK_{\Qu_i},sk_{\Qu_i})$.
Note that only those quorums linked to $\Qu_i$, and not everyone in the network,
need to know $PK_{\Qu_i}$. 
Further, every peer $p\in{}\Qu_i$ possesses a private key share $(sk_{\Qu_i})_p$ of $sk_{\Qu_i}$. 
  Unlike the quorum public/private key pair of $\Qu_i$ which
  must be known to all quorums to which $\Qu_i$ is linked in the quorum
  topology, only the members of $\Qu_i$ need to know the corresponding
  public key shares $\PK_{\Qu_i}$.
The private key $r$ of PRF $\phi (r,\cdot)$ required in \RCPqpI\ can easily be generated during this DKG execution.

\item[Routing Table Setup.]  Without loss of generality,  we assume a Chord-like DHT~\cite{stoica_etal:chord}.
When a quorum gets formed in DHTs,  it determines its neighbors and forms its routing table \RT.
For a quorum $\Qu_{i}$, each entry of its routing table has the form $\RT_{\Qu_j} = [\Qu_j, p, p', PK_{\Qu_j},$ $ts]$.
 In this entry, peer $p\in{}\Qu_j$ and peer $p'\in{}\Qu_{j-1}$ where quorum $\Qu_i$ links to quorum $\Qu_j$ and
$\Qu_{j-1}$ in the quorum topology
and $p$ and $p'$ are respectively located clockwise of all other peers in $\Qu_j$ and $\Qu_{j-1}$.
$PK_{\Qu_j}$ is the quorum public key of $\Qu_j$ generated using DKG, and $ts$ is a time stamp for when this entry was created. 
Quorum $\Qu_j$ is responsible for the identifier space between identities $p$ and $p'$.
\RT\ entries of $\Qu_{i}$ are set such that the complete identifier space is covered by them.
\end{description}

\begin{figure*}[!t]
\centering
\begin{tabular}{|m{2.6in} m{0.7in} m{2.6in}|}\hline
\multicolumn{3}{|c|}{{\bf Initial Step}: $p \in \Qu_1$ with Quorum $\Qu_1$}\\
\multicolumn{1}{|c}{\bf peer $p$} & &  \multicolumn{1}{c|}{\bf every peer $q \in  \Qu_1$}\\
&&\\
sends a request $[p|p_{\mbox{\tiny{addr}}}|ts_1]$&\centering $\Longrightarrow$&\\
& \centering $\Longleftarrow$&if the request is legitimate, reply with a signature share\\\hline

\multicolumn{3}{|c|}{{\bf Intermediate Steps}: $p \in \Qu_1$ with Quorum $\Qu_i$ for $i=2$ to $\ell-1$}\\
\multicolumn{1}{|c}{\bf peer $p$} & &  \multicolumn{1}{c|}{\bf every peer $q \in  \Qu_i$}\\&&\\
interpolate 
$\mathcal{S}_{{i-1}}=[p|p_{\mbox{\tiny{addr}}}|ts_{i-1}]_{sk_{\Qu_{i-1}}}$ using
 the received shares and send $\mathcal{S}_{{i-1}}$ and a new $ts_{i}$. Request an {\em OT initiation}&\centering $\Longrightarrow$&\\
& \centering $\Longleftarrow$&verify $\mathcal{S}_{{i-1}}$ using $PK_{\Qu_{i-1}}$ and validates $ts_i$.
If successful, send a signature share, an {\it OT-setup} message,
the {\em ranges in $\mathcal{RT}$ of ${\Qu_i}$} and the {\em entry-wise encrypted $\mathcal{RT}$ of ${\Qu_i}$}\\

interpolate 
$\mathcal{S}_{{i}}=[p|p_{\mbox{\tiny{addr}}}|ts_{i}]_{sk_{\Qu_{i}}}$ using
 the received shares and
verify it using $PK_{\Qu_{i}}$.
If invalid, sends all signature shares back.
Send an {\em OT-request for the index corresponding to the searched key}&\centering $\Longrightarrow$&\\

& \centering $\Longleftarrow$& verify all shares using $\PK_{\Qu_i}$  and inform $p$ of valid shares.
Send an {\em OT-response}\\

{\em Use the received OT-responses, if any, to determine the next quorum $\Qu_{i+1}$} &&\\\hline
 \multicolumn{3}{|c|}{{\bf Final Step}: $p \in \Qu_1$ with Quorum $\Qu_\ell$}\\
\multicolumn{1}{|c}{\bf peer $p$} & &  \multicolumn{1}{c|}{\bf $D \subseteq  \Qu_\ell$}\\&&\\
send $\mathcal{S}_{{\ell-1}}$  along with its request $m$&\centering $\Longrightarrow$&\\

\hline
\end{tabular}
\caption{\label{fig:RCP-I}\RCPqpI: \RCPI\ with Query Privacy}
\end{figure*}

\subsection{Adding Query Privacy to \RCPI: \RCPqpI}\label{sec:RCP-I}
Protocol {\sf RCP-I} works deterministically. 
Here, we include a privacy preserving mechanism for queries in \RCPI\
using the OT protocol described in Section~\ref{sec:OT}. 
The enhanced protocol (\RCPqpI) appears in Figure~\ref{fig:RCP-I}, which we outline as follows.

Assume that $p \in \Qu_1$ is searching for a \key\ and the target is a set of peers $D\subseteq{}\Qu_{\ell}$.
Let the search path go through quorums $\Qu_1, \dots, \Qu_{\ell}$. 
Peer $p$ begins by sending a request $[p, p_{\mbox{\tiny{addr}}},ts_1]$ to all
peers in its quorum $\Qu_1$., where $ts_1$ is a time stamp.
Unlike the original \RCPI, the \key\ corresponding to the intended destination of the message is not included here.
Each honest peer $q \in \Qu_1$ checks if $p$'s request follows the rule-set as described in Section~\ref{sec:Quorum}. 
If there is no violation, $q$ sends its signature share to $p$,
who interpolates those shares to generate a signature
$\Sign_{1}=[p|p_{\mbox{\tiny{addr}}}|ts_1]_{sk_{\Qu_1}}$.
In each intermediate step $(i=2$ to $\ell-1)$, $p$ sends its most recent
signature $\Sign_{i-1}$ and a new time stamp $ts_{i}$ to each peer $q \in
\Qu_{i}$. Since $\Qu_{i}$ is linked to $\Qu_{i-1}$ in
the quorum topology, each peer $q$ knows public key $PK_{\Qu_{i-1}}$ to
verify $\Sign_{i-1}$. If $\Sign_{i-1}$ is verified and $ts_{i}$ is valid, peer $q$ sends
back its signature share on $[p|p_{\mbox{\tiny{addr}}}|ts_i]$.
Peer $p$ collects the shares to form $\Sign_{{i}}$ and majority filters on the
routing information for $\Qu_{i+1}$. 
If verification of $\Sign_{{i}}$ fails, peer $p$ sends all shares back to every party in $\Qu_{i}$,
who help $p$ by filtering the invalid shares out.
Finally, for $\Qu_{\ell}$, $p$ sends $m$ along with
$\mathcal{S}_{{\ell-1}}$ to peers in the target set $D$ in $\Qu_\ell$.

\begin{figure*}[!ht]\centering
\begin{tabular}{|m{2.6in} m{0.7in} m{2.6in}|}\hline
\multicolumn{3}{|c|}{{\bf Initial Step}: $p \in \Qu_1$ with Quorum $\Qu_1$}\\
\multicolumn{1}{|c}{\bf peer $p$} & &  \multicolumn{1}{c|}{\bf every peer $q \in  \Qu_1$}\\
&&\\
sends a request $[p|p_{\mbox{\tiny{addr}}}|ts]$ &\centering $\Longrightarrow$&\\
&\centering $\Longleftarrow$&if the request is legitimate, reply with a signature share\\

verify and interpolate received shares to form
$M_1 =[p|p_{\mbox{\tiny{addr}}}|ts_1]_{sk_{\Qu_1}}$&&\\\hline

\multicolumn{3}{|c|}{{\bf Intermediate Steps}: $p \in \Qu_1$ with Quorum $\Qu_i$ for $i=2$ to $\ell-1$}\\
\multicolumn{1}{|c}{\bf peer $p$} & &  \multicolumn{1}{c|}{\bf selected peer $q_i \in \Qu_i$}\\
&&\\
select peer $q\in{}\Qu_{i}$ uniformly at random without replacement.
Send $M_{i-1}$ and request an {\em OT initiation}&\centering $\Longrightarrow$&\\

&\centering $\Longleftarrow$&For $j=i-1$ downto $2$, verify ${PK_\Qu}_{j-1}$ using ${PK_\Qu}_{j}$ and verify $M_1$
using ${PK_\Qu}_1$. If successful, send $[{PK}_{\Qu_{i-1}}]_{{sk_\Qu}_{i}}$,
the {\em ranges in $\mathcal{RT}$ of ${\Qu_i}$} and the {\em entry-wise encrypted (signed) $\mathcal{RT}$ of ${\Qu_i}$}\\

sends an {\em OT-request for the index corresponding to the searched key}&\centering $\Longrightarrow$&\\

&\centering $\Longleftarrow$&send an {\em OT-response} back\\

If ${PK_Q}_{i+1}$, $\mathcal{RT}_{\Qu_{i+1}}$ (computed from the {\it OT-response}) and ${PK_Q}_{i-1}$ verifies,
compute $M_{i} =[M_{i-1}|[{PK}_{\Qu_{i-1}}]_{{sk_Q}_{i}}]$ and
determine the next quorum $\Qu_{i+1}$ from $\mathcal{RT}_{\Qu_{i+1}}$.
Otherwise or if there is a timeout, choose $q'_i \in_R \Qu_i$ and repeat&&\\\hline

 \multicolumn{3}{|c|}{{\bf Final Step}: $p \in \Qu_1$ with Quorum $\Qu_\ell$}\\
\multicolumn{1}{|c}{\bf peer $p$} & &  \multicolumn{1}{c|}{\bf $D \subseteq  \Qu_\ell$}\\
&&\\
send $M_{{\ell-1}}$  along with its request $m$&\centering $\Longrightarrow$&\\\hline
\end{tabular}
\caption{\label{fig:RCPII}\RCPqpII: \RCPII\ with Query Privacy}
\end{figure*}

It still remains to see how $\Qu_i$ tells $p$ the correct $\Qu_{i+1}$ as the next
quorum without knowing the \key\ being searched for. We accomplish this using
the OT protocol. Along with $\Sign_{i-1}$ and $ts_i$, 
$p$ also sends an {\em OT-initiation} request to every peer in $q \in \Qu_i$.
Peer $q$ responds back with the entry-wise symmetrically 
encrypted (AES) routing table $\RT_{\Qu_i}$, the {\it OT-setup} message, and the upper and lower bounds of ranges in $\RT_{\Qu_i}$.
Note that since all quorum members use the same randomness (due to the use of a PRF where everyone holds the private key), 
the messages from all honest parties will be the same.
Peer $p$ determines an index in $\RT_{\Qu_i}$ for the next quorum by searching for \key\ in the received ranges
and sends an {\em OT-request} for that index. Peer $q$ then computes and
sends the {\em OT-response}.
Using this response, peer $p$ obtains the symmetric key corresponding to the queried index
and decrypts the appropriate entry in  $\RT_{\Qu_i}$ to determine the next quorum $\Qu_{i+1}$.
Any wrongdoing by Byzantine peers in $\RT_{\Qu_i}$ range tables, encrypted $\RT_{\Qu_i}$ blocks 
and OT executions are taken care of by the majority action.
As $p$ knows $\Qu_2$ using its own routing table, there is no OT involved in the initial step.

Notice that it is also possible for peer $p$ to use OT in the final step
while communicating with the target set $D$ in $\Qu_\ell$, if the privacy application demands it.
In that case, the target set $D$ only knows that the queried key is one of its keys, but
cannot determine the exact key.

The correctness of protocol \RCPqpI\ follows directly from that of protocol {\sf RCP-I}
and we refer the readers to \cite{ToN} for a detailed proof. 
Although the  encrypted routing tables (a few kilobytes in size) are sent in our privacy-preserving approach 
as compared to the individual routing table entires in \RCPqpI, it does not affect the message complexity of the protocol.
The message complexity of protocol \RCPqpI\ remains exactly the same as protocol \RCPI,
which is equal to $O(\log^2{n})$.
We discuss the increase in computational cost and other systems matters in Section~\ref{sec:System}.

\subsection{Adding Query Privacy to \RCPII: \RCPqpII} \label{sec:RCP-II}
Protocol {\sf RCP-II} utilizes signed routing table (\RT) information and 
reduces the message complexity in protocol {\sf RCP-I} by a linear factor (in expectation)
using a uniformly random selection of peers in the quorums.
Here, all \RT\ entries are signed separately by the quorum whenever
\RT{}s are modified.
In particular, every peer in the quorum,  using their DKG private key shares, 
generates and sends signature shares, which are then interpolated to
obtain signed \RT\ entries.
The OT setup and the OT protocol remain exactly the same as in \RCPqpI.
The enhanced protocol (\RCPqpII) appears in Figure~\ref{fig:RCPII}, which we outline as follows.

Initially, for simplicity, assume that peers act correctly.
The initial step, where $p$ communicates within its quorum, $\Qu_1$,
remains exactly the same. Each peer in $\Qu_1$ receives $[p | p_{\mbox{\tiny{addr}}} | ts]$
from $p$. 
If the request does not violate the rule set, then peer $p$ receives signature shares and computes
$M_1=[p, p_{\mbox{\tiny{addr}}}, ts]_{{sk_\Qu}_1}$.
Next, $p$ knows the membership of $\Qu_{2}$ which belongs to its \RT,
and selects a peer $q_2\in{\Qu_2}$ uniformly at random without replacement.
Peer $p$ sends $M_1$ to $q_2$.  The correct $q_2$ verifies $M_{1}$ using ${PK_\Qu}_1$, 
and replies with $[{PK_\Qu}_1]_{{sk_\Qu}_2}$ and
$[\mathcal{RT}_{\Qu_3}]_{{sk_\Qu}_2}$. 
Here, $[{PK_\Qu}_j]_{{sk_\Qu}_i}$ denotes the quorum public key of $\Qu_j$ signed by quorum $\Qu_i$ as 
neighboring quorums know each others' public keys, and
$[\mathcal{RT}_{\Qu_j}]_{sk_{\Qu_i}}$ denotes the routing entry for $\Qu_j$ signed by $\Qu_i$.
Peer $p$ verifies $[{PK_\Qu}_1]_{{sk_\Qu}_2}$ and $[\mathcal{RT}_{\Qu_3}]_{{sk_\Qu}_2}$, 
and checks if the time stamp is valid.
If so, $p$ constructs $M_2=[M_1 | [{PK_\Qu}_1]_{{sk_\Qu}_2}]$.
The idea is to allow some peer in $\Qu_3$ to verify ${PK_\Qu}_1$ and $M_1$
using a signature chain. Further,  $p$ can check the response from some peer in $\Qu_3$ in the next step
using ${PK_\Qu}_3$ included in $\mathcal{RT}_{\Qu_3}$.
This process repeats with minor changes for the remaining steps
until $p$ reaches the destination quorum $\Qu_\ell$.
If any peer does not respond in the amount of time predefined by the weak synchrony assumption~\cite{castro:practical}
(as described in Section~\ref{sec:Assump})
or responds incorrectly, the protocol proceeds by choosing uniformly at random another peer in the quorum.
Note that any attempt by a malicious peer to return incorrect information is detectable.

It still remains to see how the OT executions for \key\ are performed such that a correct peer $q_i$ in $\Qu_i$
can give routing information for $\Qu_{i+1}$ to peer $p$. For this, peer $p$ sends an {\em OT initiation}
to peer $q_i$ along with $M_{i-1}$.  Upon verification of the signature chain, $q_i$ replies with
$[{PK}_{\Qu_{i-1}}]_{{sk_\Qu}_{i}}$,  ranges in $\mathcal{RT}$ of ${\Qu_i}$, 
entry-wise encrypted (signed) $\mathcal{RT}$ of ${\Qu_i}$, and the OT-setup message.
Note that these encryptions are done locally at peers, and applied on both the \RT\ entires and signatures.
Peer $p$ then determines an index corresponding to the \key\ it is searching for and
sends an {\em OT-request} for that index. Peer $q_i$ then computes and sends an {\em OT-response}.
Using this response, peer $p$ obtains the symmetric key corresponding to the queried index
and decrypts the appropriate entry in  $\RT_{\Qu_i}$, checks the
signature on the resulting plaintext, and thus determines the next quorum $\Qu_{i+1}$.

Similar to \RCPqpI, if required, it is possible for peer $p$ to use OT in the final step
while communicating with the target set $D$ in $\Qu_\ell$.
The correctness of the protocol follows directly from that of the
original \RCPII\ protocol,
and we refer the readers to \cite{ToN} for a detailed proof. 
The message complexity of the enhanced protocol remains exactly the same as the original protocol,
which is equal to $O(\log{n})$ in expectation.
We discuss the increase in computational cost and other systems matters in Section~\ref{sec:System}.

\section{Analysis and Discussion}\label{sec:System}
As discussed in Section~\ref{sec:schemes}, our protocols do not increase the message complexity of
their original counterparts \RCPI\ and \RCPII. 
In this section, we consider the increase in computation due to the query-privacy mechanism
and find it to be nominal.
We also analyze possible system-level attacks on our protocols. 

\subsection{Additions to Computational Costs}
Query privacy does not come without some additional computation.
However, for our choice of OT, this increase is insignificant as compared to the computations already 
done in the original \RCPI\ and \RCPII\ protocols.

In both the \RCPqpI\ and \RCPqpII\ protocols, a requesting peer $p$ has to perform only two additional exponentiations 
at each privacy-preserving \RT\ entry retrieval, 
while a responding  peer $q_i$ in quorum $\Qu_i$ must perform one additional exponentiation.
Peers in $\Qu_i$ also have to perform $\nu$ exponentiations for an OT setup,
where $\nu$ is the size of \RT.
However, they can be batch-computed and may also be reused in $\nu$ requests.
In terms of computation, our privacy-preserving mechanism remains exactly the same in both protocols, \RCPqpI\ and \RCPqpII.
This results from a peer $p$ running the same instance of OT with all peers in the quorum in \RCPqpI\
with the help of the PRF-based technique discussed in Section~\ref{sec:OT}.

Timing values computed using the pairing-based cryptography (PBC) library~\cite{pbclibrary} 
indicate that one exponentiation takes around $1$ ms on a desktop machine. Given that the
communication time for the original \RCPI\ and \RCPII~protocols is greater than $3$ seconds
(refer to Young et al.~\cite{young:practical} for a detailed discussion), the cost of these exponentiations is insignificant.
In terms of system load, a DKG execution in \RCPI\ and \RCPII\ on average requires $2$ CPU seconds,
and a threshold signature generation and verification takes about $6$ CPU ms. Therefore, our OT
executions do not increase the system load by any significant fraction.
Note that the OT protocol also involves a few group multiplications, PRF executions,
symmetric encryptions and hashes. 
Their computations take only a few $\mu s$, so we ignore these computational costs in our discussion.

\subsection{System-level Attacks on Query Privacy}
Although OT hides the queried \key\ completely in the cryptographic sense,
there can be system-level attacks that leak some information about the \key.

A range estimation attack defined by Wang, Mittal and Borisov~\cite{WMB10}
that reduces privacy provided in NISAN~\cite{PRR09} could be applied to our \RCPqpI\ protocol.
This attack is based on the fact that the Chord-like DHT ring is directed and
the requesting peer $p$ will not query a quorum succeeding the queried \key\ except in the first iteration.
Therefore, an adversary that can observe the peer $p$ contacting a 
sequence of quorums can put them together into a 
sequence to narrow down on the target range that peer $p$ may reach.
In this attack, the range only extends from the last contacted quorum having 
an adversarial peer to the largest jump possible at the end of first iteration.
For NISAN, Wang et al.\ show that if at least $20\%$ of nodes are under the adversary's control, 
the adversary may obtain a significant amount of information about the queried key. 
As indicated in Section~\ref{sec:Assump}, 
we consider the percentage of peers under the control of a single adversary to be around  $10\%$. 
Therefore, although this range estimation attack is  possible, it is not particularly effective 
in our DHT setting.
On the other hand, the curious peers in the intermediate quorums only see 
requests approved by one of their neighbors.
This, along with the security provided by the OT protocol,
ensures that  nothing is revealed about the queried \key\ to the curious intermediate quorums.

As only an expected constant number of peers are contacted per intermediate quorum in our \RCPqpII\ protocol,
the range estimation attack by Wang et al.\ \cite{WMB10} is far less effective.
However, query privacy for our \RCPqpII\ protocol is slightly weaker in terms of the above mentioned
curious observer attack.
This is a direct consequence of the use of a signature chain to authorize a request from a peer $p$:
assume a peer $q_i$ from an intermediate quorum $\Qu_i$.
Although $q_i$ may not be able to determine quorums from the public keys in a chain,
the length of the chain itself might give peer $q_i$ some information about possible \key{} values.
This results from a property of Chord-like DHTs: {\em generally}
each step brings a requester exponentially closer to its destination.
As an example, a shorter signature chain indicates that 
a destination quorum is probably situated away from $\Qu_i$ in the key (or identifier) space, 
while a length nearly equal to $\log n$ indicates that
 the destination quorum is probably nearby.
This is, however, a weak heuristic attack as path lengths of DHT requests may vary significantly.
Further, it is possible to mislead such a curious adversary by adding a few fake signatures
at the end of the chain. The requesting peer $p$ has to have this done by its quorum $\Qu_1$.

\subsection{Crawling Attacks towards Spam Prevention}
 As discussed in Section~\ref{sec:Related},
usage of the iterative routing approach significantly improves robustness against spamming attacks, since
a spamming peer has to perform an equal amount of work as the rest of the system.
Young et al.\ \cite{young:practical} add further protections against spamming in \RCPI\
and \RCPII\ by not allowing the adversarial peer to gather a large amount of routing information.
They add the queried keys to requests. As a result, an execution of \RCPI\ or \RCPII\
leads to the requester $p$ gaining information only about the $\ell$ quorums in its path.
We concentrate on query privacy in this work and enforce that
the queried \key\ should remain completely oblivious to every intermediate quorum $\Qu_i$ for $i \in [1,\ell-1]$.
This may lead to attacks, where the adversary peer $p$ obtains more routing information;
we call these attacks {\em crawling attacks}. 

In our \RCPqpI\ protocol, 
a malicious peer $p$ may try to obtain the entire \RT\ of $\Qu_i$ by querying for
different \key s (or \RT\ indices) to different peers in $\Qu_i$. 
As a result, the adversary peer $p$ can acquire more information than allowed by
the rule set.
It is possible to thwart this supposed attack completely by adding one communication round: here,
$p$ also has to get its {\em OT-request} message (which is the same for all peers in $\Qu_i$) signed from $\Qu_i$
in the exact same way as its  authorization request $[p|p_{\mbox{\tiny{addr}}}|ts_i]$.
This ensures that $p$ can query the quorum for only one \key\ (specifically, one index in \RT),
and query privacy of the \key\ also remains unaffected. 
This additional one round does not change the message complexity of the protocol.
We do not include this defense mechanism in the protocol described in Figure~\ref{fig:RCP-I},
as repercussions of this attack, if any, may vary from system to system.

In our \RCPqpII\ protocol,
similar crawling is possible. The adversary peer $p$ may query different peers in quorum
$\Qu_i$ for different indices to obtain the complete \RT\ for $\Qu_i$.
However, unlike in $\RCPqpI$, a malicious peer $p$ has to increase its effort linearly
to obtain the complete \RT\ of $\Qu_i$ in \RCPqpII\ and crawling is not an effective attack
for the malicious peer $p$.

In both protocols, it is possible for a malicious peer $p$ to alter the queried \key\
while shifting from one quorum to the next, as
there is no link between signed authorizations and the queried \key s for privacy reasons. 
This may, however, lead to a peer $p$ gaining more knowledge as
it can continuously modify its \key\ to traverse as much of the DHT as possible.
This is an even weaker crawling attack than the one mentioned above,
as the adversary has to perform a significant amount of work to gain any information.

Notice that any information gained by the adversary in the above active attacks
is still substantially smaller than information effortlessly available to it when PIR or trivial PIR are used.
Finally, it may be possible to stop the adversary $p$ from gaining any additional information
without revealing its key using computationally and communicationally
demanding cryptographic primitives such as zero-knowledge proofs or conditional OT~\cite{DOR99}.
However, we find that their inclusions are not essential, and may be even impractical, for DHT-based systems.

\section{Conclusion}\label{sec:Con}
In this paper, we have introduced the concept of query privacy in the robust DHT architecture.
We have enhanced two existing robust communication protocols (\RCPI\
and \RCPII) over DHTs
to preserve the privacy of keys in DHT queries using an OT protocol.
We reviewed the OT literature and chose a theoretically non-optimal but practically efficient 
(in terms of use over DHTs in practice) OT scheme.
Using this, we built two protocols
(\RCPqpI\ and \RCPqpII), which
obtain query privacy without any significant increase in computation costs and message complexity in practice.
Our privacy-preserving mechanism
does not change the underlying protocols' utility or efficacy in any way, and
is also be applicable to other  DHT communication architectures.

\bibliographystyle{abbrv}
\bibliography{anonymous-p2p}
\appendix
\section{The Oblivious Transfer Protocol}\label{app:OT}
In this appendix, we
provide an overview of the \ooom{\nu} OT protocol of Naor and Pinkas
\cite{NP01}. Security of the construction is based on DDH in a group
\OTG{} of prime order $|\OTG|$. The proof of security uses the random
oracle model, i.e.~the protocol uses a cryptographic hash function,
$H$, which is then replaced by the random oracle in the proof. Recall
that the goal is for the server, $q$, to offer $\nu$ strings,
$S_1,\ldots, S_\nu$, and for chooser $p$ to obtain the desired one,
$S_\rho$, and nothing else. The basic idea of the \cite{NP01} OT
protocol is to let $p$ provide encryption keys $\OTPK_i$ for $1\leq i
\leq \nu$; these are constructed such that $p$ can know at most one of
the decryption keys.  The server then supplies $p$ with encryptions of
each $S_i$ under $\OTPK_i$. Details are included in a protocol flow below in Figure~\ref{fig:OT}.
we now elaborate on the intuition behind the three messages:
\begin{enumerate}
\item {\bf \textit{OT-setup}}: $q$ picks a random DL instance,
  $\alpha=g^r$ and sends this to $p$. Moreover, the parties agree on
  $\nu-1$ random group elements, $C_2,\ldots,C_\nu$. It is crucial
  that $p$ does not know their DL, hence they are picked by $q$ (who
  is allowed but not required to know the DL).
\item {\bf \textit{OT-request}}: $p$ will supplies $\OTPK_1$ to $q$;
  for $1<i \leq \nu$ $\OTPK_i$ is implicitly set to
  $C_i/\OTPK_1$. $p$ constructs $\OTPK_1$ such that $\OTPK_\rho$ has
  known DL; however, if $p$ could find the DL of any other key,
  $\OTPK_i$, $p$ could solve a DDH problem in \OTG{}.
\item {\bf \textit{OT-response}}: $q$ computes $\OTPK_i^r =
  C_i^r/PK_1^r$ for all $i$. Since the $C_i^r$ may be precomputed
  this requires only a single exponentiation and $\nu-1$
  multiplications. $q$ then picks a uniformly random $\ell$-bit string,
  $R$, where $\ell$ is chosen large enough (e.g.~200 bits) to ensure
  that $R$ will be distinct. Finally, for each $1\leq i \leq \nu$ $q$
  computes an encryption of $S_i$ as $E_i = H(\OTPK_i^r, R, i) \oplus
  S_i$.  $R$ and the $E_i$ are sent to $p$, who computes first
  $\OTPK_\rho^r = \alpha^k$ and then decrypts to obtain $S_\rho = E_\rho \oplus
  H(\OTPK_\rho^r, R, \rho)$.
\end{enumerate}
For more details along with the proof in the random oracle model, see
\cite{NP01}. As noted, $\alpha$ and the $C_i^r$ may be preprocessed
during periods of low computational load. Moreover, the values may be
used in multiple instances of the OT protocol. ``Refreshing'' $r$
(i.e.~$\alpha$ and the $C_i^r$) every $\nu$ execution provides an
amortized complexity of two exponentiations per party per OT
invocation. The setup message consists of $\nu$ group elements, while
the OT-request contains only a single one. Finally, the reply consists
of $\nu$ $E_i$s plus $R$; though strictly speaking, these are not
group elements, they may be viewed as such for the complexity
analysis. Hence, overall communications is $2\nu + 2$ group elements.

We
remark that since we are transferring AES keys using OT, we could also
directly use (some digest of) $\OTPK_i^r, R, i$ as the AES
key. However, such ad hoc optimizations may easily introduce subtle
flaws. The security proof of Naor and Pinkas may easily be invalidated
by even a minor optimization, hence, as the gains are marginal, we
prefer the original OT protocol to any ad hoc optimization.

\begin{figure*}[!t]
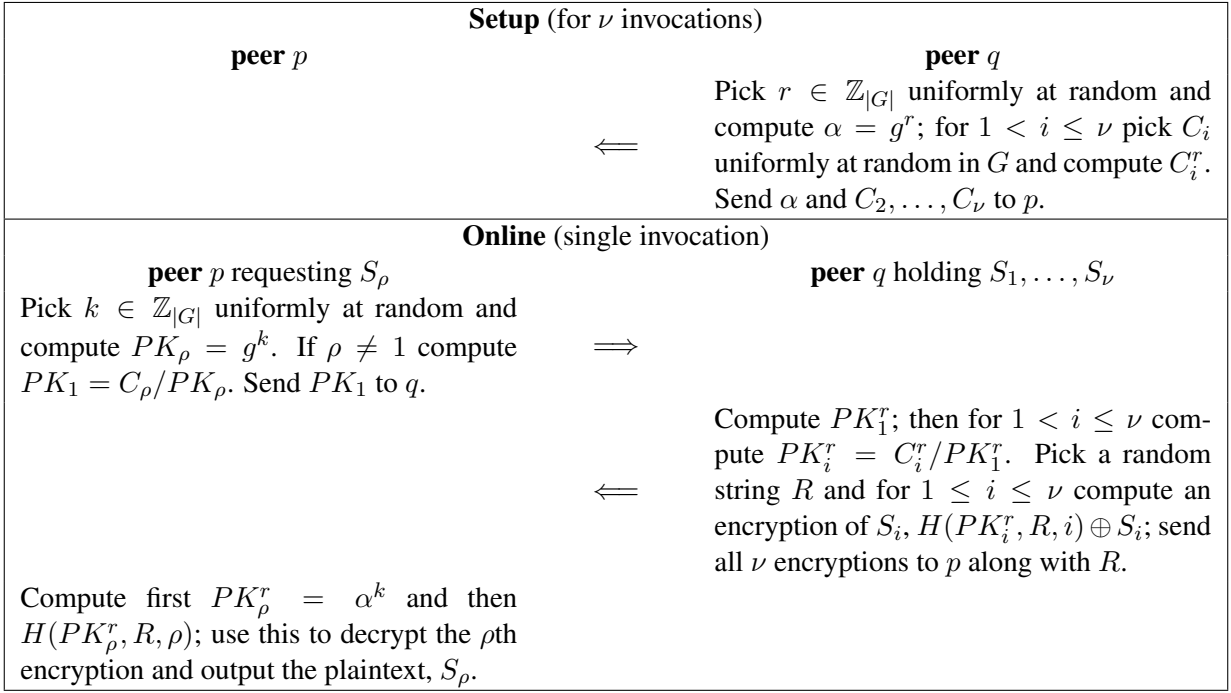

\centering
\begin{tabular}{|m{2.6in} m{0.7in} m{2.6in}|}\hline 
\multicolumn{3}{|c|}{{\bf  Setup} (for $\nu$ invocations)}\\
\multicolumn{1}{|c}{\bf peer $p$} & &  \multicolumn{1}{c|}{\bf peer $q$}\\
&\centering $\Longleftarrow$&Pick $r\in\ZOTG$ uniformly at random and compute
$\alpha = g^r$; for $1<i\leq \nu$ pick $C_i$ uniformly at random in
\OTG{} and compute $C_i^r$. Send $\alpha$ and $C_2,\ldots,C_\nu$ to $p$.
\\\hline
\multicolumn{3}{|c|}{{\bf Online} (single invocation)}\\ \multicolumn{1}{|c}{{\bf peer
    $p$} requesting $S_\rho$} & & \multicolumn{1}{c|}{{\bf peer $q$}
  holding $S_1,\ldots,S_\nu$}\\ Pick $k\in \ZOTG$ uniformly at random
and compute $\OTPK_\rho = g^k$. If $\rho \neq 1$ compute $\OTPK_1 =
C_\rho/\OTPK_\rho$. Send $\OTPK_1$ to $q$.&\centering$\Longrightarrow$&\\

& \centering$\Longleftarrow$& Compute $\OTPK_1^r$; then for $1<i\leq \nu$
compute $\OTPK_i^r = C_i^r/\OTPK_1^r$.  Pick a random string $R$ and
for $1\leq i\leq \nu$ compute an encryption of $S_i$, $H(\OTPK_i^r,
R, i)\oplus S_i$; send all $\nu$ encryptions to $p$ along with
$R$.\\ {Compute first $\OTPK_\rho^r = \alpha^k$ and then $H(\OTPK_\rho^r, R,
  \rho)$; use this to decrypt the $\rho$th encryption and output the
  plaintext, $S_\rho$.} &&\\\hline
\end{tabular}
\caption{\label{fig:OT} The \ooom{\nu}~OT protocol of Naor and Pinkas}
\end{figure*}


\end{document}